\documentclass[%
 aip,
 apl,%
 amsmath,amssymb,
 reprint,%
]{revtex4-1}

\usepackage{graphicx}
\usepackage{dcolumn}
\usepackage{bm}

\begin{document}

\title{Amplicification of Voltage Controlled Magnetic Anisotropy Effect with Negative Capacitance}

\author{Lang Zeng}
\email{zenglang@buaa.edu.cn.}
\author{Tianqi Gao}
\thanks{Lang Zeng and Tianqi Gao contributed equally to this work therefore they are co-first authors.}
\author{Deming Zhang}
\author{Shouzhong Peng}
\affiliation{Spintronics Interdisciplinary Center, Beihang University, Beijing, 100191 China}
\affiliation{School of Electrical and Information Engineering, Beihang University, Beijing, 100191 China}
\author{Fanghui Gong}
\affiliation{Spintronics Interdisciplinary Center, Beihang University, Beijing, 100191 China}
\affiliation{School of Mathematics and Systems Science, Beihang University, Beijing, 100191, China}
\author{Xiaowan Qin}
\author{Mingzhi Long}
\author{Youguang Zhang}
\author{Weisheng Zhao}
\email{weisheng.zhao@buaa.edu.cn}
\affiliation{Spintronics Interdisciplinary Center, Beihang University, Beijing, 100191 China}
\affiliation{School of Electrical and Information Engineering, Beihang University, Beijing, 100191 China}

\date{\today}

\begin{abstract}
The high current density ($10^6\sim 10^8~A/cm^2$) required by Magnetic Tunneling Junction (MTJ) switching driven by Spin Transfer Torque (STT) effect leads to large power consumption and severe reliability issues therefore hinder the timetable for STT Magnetic Random Access Memory (STT-MRAM) to mass market. By utilizing Voltage Controlled Magnetic Anisotropy (VCMA) effect, the MTJ can be switched by voltage effect and is postulated to achieve ultra-low power (fJ). However, the VCMA coefficient measured in experiments is far too small for MTJ dimension below 100 nm. Here in this work, a novel approach for the amplification of VCMA effect which borrow ideas from negative capacitance is proposed. The feasibility of the proposal is proved by physical simulation and in-depth analysis.
\end{abstract}

\maketitle

Magnetic Tunneling Junction (MTJ) is the most important and fundamental building blocks in spintronics devices~\cite{MTJ1,MTJ2,MTJ3}. Combining with Spin Transfer Torque (STT) effect~\cite{STT1,STT2}, the STT Magnetic Random Access Memory (STT-MRAM) is considered to be a very promising candidate for the next generation of universal memory due to its large tunneling magnetoresistance ratio (TMR), current control of magnetization switch, non-volitality and fast switch~\cite{MRAM1,MRAM2,MRAM3}. MTJ structure is also widely employed as building blocks to generate spin polarized current or electrical readout in spintronics logic devices~\cite{spinlogic1,spinlogic3}. In neuromorphic computation, spintronics synapses and neurons are proposed based on MTJ~\cite{neural1,neural4}. However, the current density required by STT effect to switch free layer of MTJ is in the range of $10^6\sim 10^8~A/cm^2$. Such high current density will cause high power consumption and severe reliability issues. The situation will be even worse when further scaling of STT-MRAM enters into sub-volume region($<40~nm$)~\cite{MRAM1,MRAM2,MRAM3}.

A number of proposals has been set forth for the reduction of switching current and delay. Most of the proposed approaches are based on the observation of existing of incubation period when STT-MRAM switches~\cite{SHE1,orthogonal1}. The method to eliminate incubation time is to utilize an additional orthogonal torque. This orthogonal torque can be generated in generally two ways: 1) Injecting charge current through a magnetic layer with orthogonal magnetization~\cite{orthogonal1}. 2) Spin Hall Effect (SHE)~\cite{SHE1}. Among them, novel MTJ structure with SHE effect is a hot and fast-progress research field to show low switch current density and small switch delay. Besides these, there is another approach based on electrical field effect which is called Voltage Controlled Magnetic Anisotropy (VCMA) effect~\cite{VCMA1,VCMA2,VCMA3}. The Perpendicular Magnetic Anisotropy (PMA) that is crucial for STT-MRAM scaling into sub-volume region ($< 40~nm$) originates from the interfacial effect between MgO and CoFeB~\cite{MTJ3}. PMA is enhanced by the change of atomic orbit occupation following electron migrates across the interface~\cite{PMA1,PMA2}. It is natural to deduce that applying electrical field also will cause electron transfer between MgO and CoFeB therefore can enhance or weaken PMA.

The most significant feature of VCMA effect is it can convert the current driven STT-MRAM into a voltage driven device. As we all know, voltage driven devices intrinsically has lower energy consumption comparing with current driven devices. In addition, the leakage current through the tunneling barrier in VCMA driven MTJ can be negligible since much thicker insulator can be used. The severe reliability issues accompanied by high switching current density will be relieved. However, as MTJ dimension shrinks, the PMA should become stronger to keep constant $\Delta E$ with smaller volume of magnetic layer. This will demand very large VCMA coefficient. For MTJ free layer dimension of $20~nm\times 20~nm\times 1~nm$, the effective perpendicular anisotropy $K_{eff}$ must be higher than $3.2946\times 10^{11} A^2/m^2$ to keep a constant $\Delta E$ at $40~kT$. This will demand a VCMA coefficient as high as $414~fJ/V\cdot m$ at VCMA voltage of $1~V$ and MgO thickness of 1 nm to fully suppress the interfacial PMA. However, as summarized in Table.~\ref{table1}, the maximum experimental reachable VCMA coefficient is only $\sim 100~fJ/V\cdot m$ which is far too small. Small VCMA coefficient is the biggest obstacle for voltage driven ultra-low power (fJ) MRAM in sub-volume region.

\begin{table*}
\caption{\label{table1}Experimental Measured VCMA Coefficient for Different Seed/Cap Layer and Free Layer}
\begin{ruledtabular}
\begin{tabular}{cccccccc}
Seed/Cap Layer &Ta&Ta&MgO&Au&Ru&Hf&Mo\\
\hline
Free Layer&CoFeB&Fe/FeB&FeB&CoFe&CoFeB&CoFeB&CoFeB\\
VCMA Coefficient ($fJ/V\cdot m$)&$30-60~\cite{VCMA6,VCMA7}$&$10\cite{VCMA8}$&$108~\cite{VCMA9}$&$30-40~\cite{VCMA11}$&$18~\cite{VCMA10}$&$10~\cite{VCMA13}$&$40~\cite{VCMA12}$\\
\end{tabular}
\end{ruledtabular}
\end{table*}

As shown in Table.~\ref{table1}, people make different combinations of seed/cap layer and free layer trying to increase VCMA coefficient~\cite{VCMA6,VCMA7,VCMA8,VCMA9,VCMA10,VCMA11,VCMA12,VCMA13}. The interface between MgO and CoFeB can be further polished to increase VCMA coefficient. Here in this work, we propose an alternative but novel approach to amplify VCMA effect. The idea is borrowed from negative capacitance which is proposed for low sub-threshold slope CMOS device~\cite{NC1}. The voltage across MgO is amplified by negative capacitance thus a much larger effective VCMA effect can be achieved. The negative capacitance is realized by Ferroelectric (FE) material and physical simulation of FE layer under Landau theory and Landau-Khalatnikov equation is performed. The total capacitance of Dielectric-FE (DE-FE) double layer is also calculated to show the feasibility of the proposal approach.

\begin{figure}
\includegraphics[width=0.4\textwidth]{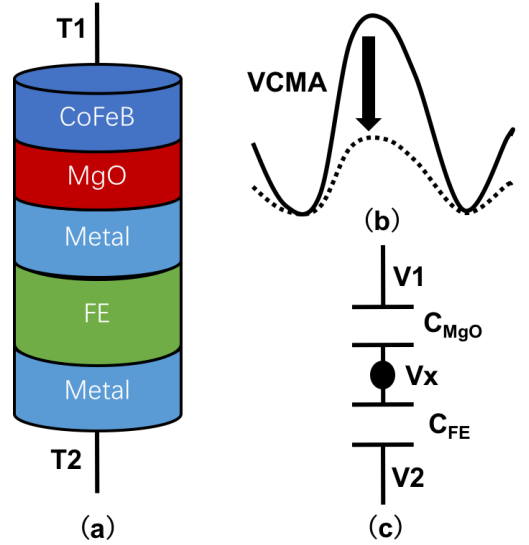}
\caption{\label{fig1} (a) The structure proposed in this work which can amplify VCMA effect by negative capacitance. (b) The skematic illustration of working machanism of VCMA effect. (c) The simplied capacitance model for amplification of VCMA effect.}
\end{figure}

The proposed structure which can utilize negative capacitance to amplify VCMA effect is shown in Fig.~\ref{fig1}a. The CoFeB layer is the magnetic layer whose magnetic anisotropy can be controlled by applied electrical field. The MgO layer is insulating layer and is lattice matched with CoFeB magnetic layer. The other reason to choose MgO is that MgO can enhance the interfacial PMA of CoFeB layer~\cite{PMA1,PMA2}. Applying a positive (negative) voltage across the MgO reduces (increases) PMA of CoFeB layer.The metal layer between MgO and FE layer is to relax the possible strain caused by lattice mismatch. The negative capacitance is realized by FE layer which we will simulate and analyze in the next paragraph. The bottom metal layer is for contact. The simplified capacitance model for the amplification of VCMA effect is shown in Fig.~\ref{fig1}c. For $V_1 = 0$, the voltage $V_x$ can be written as
\begin{equation}
V_x = \frac{V_2 C_{FE}}{C_{MgO}+C_{FE}}
\end{equation}
Assume $C_{MgO}+C_{FE}\approx 0$, the voltage across MgO $|V_x|$ can be much higher than $V_2$. So the voltage across MgO is amplified leading to a much larger VCMA effect. The total capacitance amplification ratio is defined as
\begin{equation}
\frac{C_{tot}}{C_{MgO}} = \frac{C_{FE}}{C_{MgO}+C_{FE}}
\label{Ctot}
\end{equation}
It is easily found that the magnitude of voltage amplification can be calculated by the magnitude of capacitance amplification.

\begin{figure}
\includegraphics[width=0.45\textwidth]{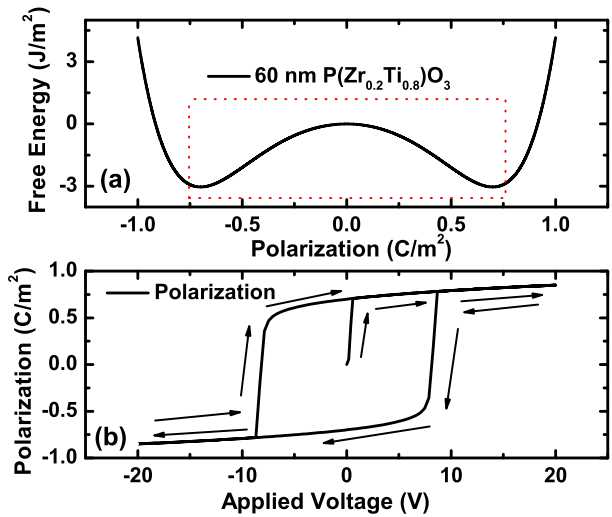}
\caption{\label{fig2} (a) The free energy landscape of ferroelectic material P($\rm{Zr_{0.2}Ti_{0.8}}$)$\rm{O_3}$ (PZT) simulated with Landau theory. The red dot box indicates the negative capacitance region. (b) The Charge (or polarization)-Voltage characteristics of a ferroelectric material PZT simulated by Landau-Khalatnikov equation.}
\end{figure}

The negative capacitance realized by ferroelectric material can be explained with Landau theory~\cite{landau1}. In terms of free energy, the capacitance can be defined as follows
\begin{equation}
C = \frac{d^2U}{dQ^2}
\label{capacitance}
\end{equation}
where $U$ is the free energy, and $Q$ is charge or polarization. It is obvious that negative curvature region in free energy landscape of an insulating material corresponds to a negative capacitance. The free energy can be describe by Landau theory with an order parameter. In the case of ferroelectric (dielectric) material, this order parameter is the polarization (charge). In Landau theory, the free energy of $U$ is an even order polynomial of the polarization $Q$, written as
\begin{equation}
U = d\cdot(\alpha_1 Q^2+\alpha_{11} Q^4+\alpha_{111} Q^6)-VQ
\end{equation}
Here, $d$ is the ferroelectric material thickness, $V$ is the voltage applied across the ferroelectric layer, $\alpha_{1}$, $\alpha_{11}$ and $\alpha_{111}$ are rescaled anisotropy constants. It is worth noting that $\alpha_{11}$ and $\alpha_{111}$ are temperature independent; $\alpha_{1}=\alpha_0(T-T_0)$ where $\alpha_0$ is a temperature independent positive quantity, $T$ and $T_C$ are the temperature and the Curie temperature respectively. It is can be seen that below Curie temperature $\alpha_1 < 0$ which leads to the negative curvature of the free energy landscape of a ferroelectric layer when $Q\approx 0$. For instance, we have shown the free energy landscape of a 60 nm thick P($\rm{Zr_{0.2}Ti_{0.8}}$)$\rm{O_3}$ (PZT) in Fig.~\ref{fig2}a. It can be observed that when the polarization $|Q| < 0.7~C/m^2$, the free energy landscape is negative curved thus corresponding to negative capacitance. The values of the rescaled anisotropy constants for ferroelectric PZT are taken from Rabe's~\cite{NC3}.

The unstable of negative capacitance can be inferred from Fig.~\ref{fig2}a since the second derivative $d^2U/dQ^2 < 0$. The instability of negative capacitance also can be understood from the view point of polarization-voltage hysteresis loop. The dynamic hysteresis loop of ferroelectric material is modeled in accordance with Landau-Khalatnikov equation which can be written as
\begin{equation}
\rho \frac{dQ}{dt} = -\frac{dU}{dQ}
\end{equation}
where $\rho$ accounting for dissipative processes during ferroelectric switching is a material dependent parameter in the units of $\Omega$. The quasi-static calculation of ferroelectric switch governing by Landau-Khalatnikov equation for a 60 nm thick PZT is shown in Fig.~\ref{fig2}b. Starting from $V=0$ and $Q=0$, the ferroelectric PZT layer quickly gets spontaneous polarization around $0.7~C/m^2$ and deviates from the negative capacitance region.

\begin{figure}
\includegraphics[width=0.45\textwidth]{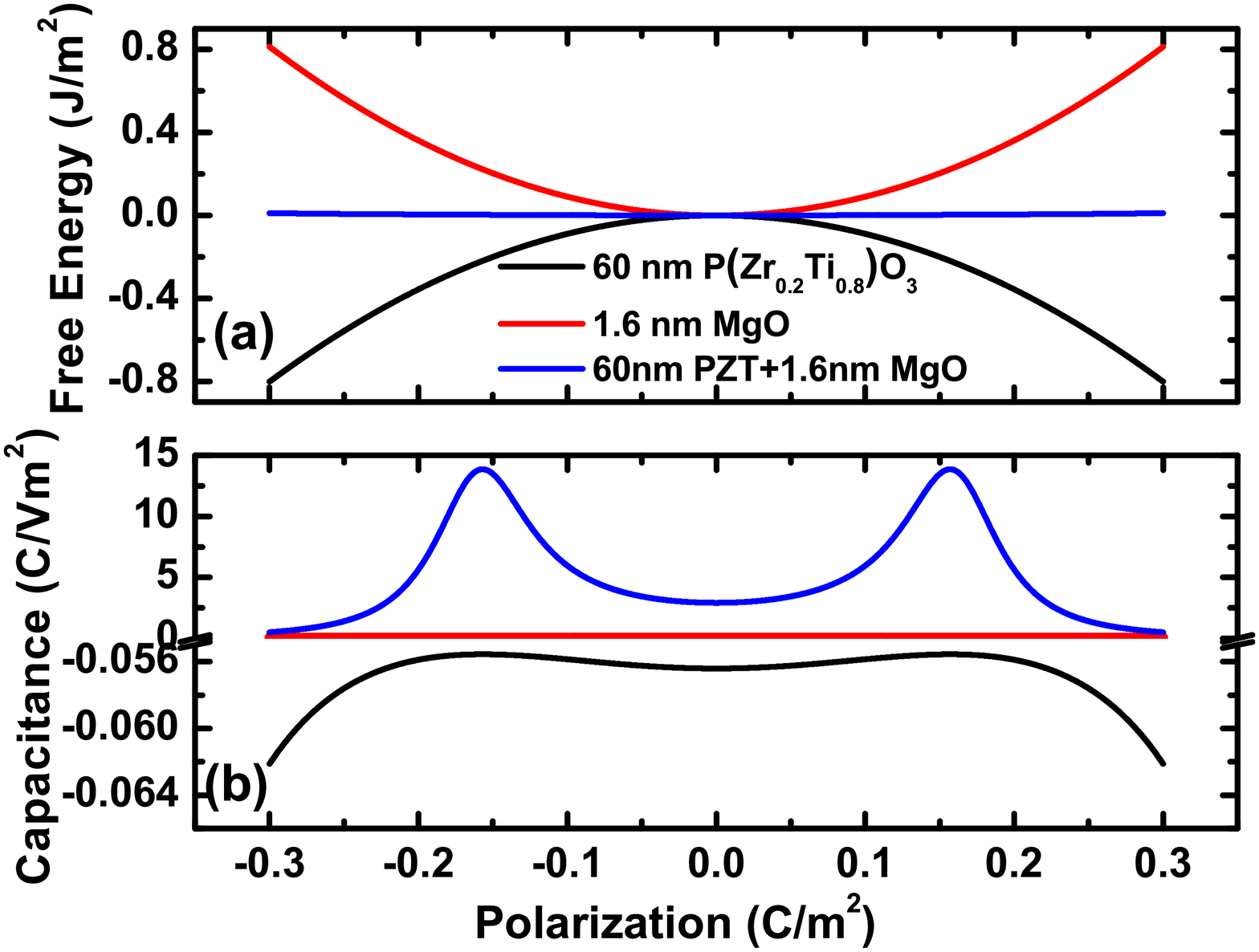}
\caption{\label{fig3} (a) The free energy landscape of FE layer, DE layer and FE-DE double layer. Since capacitance can be defined as $C = d^2U/dQ^2$, the flatter the energy landscape is, the larger the capacitance is. (b) The capacitance of FE layer, DE layer and FE-DE double layer. The amplification of total capacitance is shown.}
\end{figure}

The ferroelectric negative capacitance can be stabilized by a Dielectric-Ferroelectric (DE-FE) double layer which was proposed by Prof. Datta~\cite{NC1}. As shown in Fig.~\ref{fig1}c and Eq.~\ref{Ctot}, the total capacitance $C_{tot} = (C^{-1}_{FE}+C^{-1}_{MgO})^{-1}$ will be positive if $C_{FE}+C_{MgO}<0$ when $C_{FE}<0$. So by connecting a positive capacitance with appropriate value in series with a negative capacitance, the ferroelectric material can be pinned and stabilized in negative capacitance region. The stabilization of negative capacitance and total capacitance amplification can be simulated with Landau theory. The total free energy of DE-FE double layer $U_{DE+FE}$ is the sum of the free energies of the ferroelectric layer $U_{FE}$ and the dielectric layer $U_{DE}$ respectively. The free energy landscape of 60 nm PZT layer, 1.6 nm MgO and PZT-MgO double layer are shown in Fig.~\ref{fig3}a. The dielectric constant of MgO is chosen to be $10$. The thickness of MgO layer is 1.6 nm which is the appropriate value for the stabilization of the negative capacitance. As shown in Fig.~\ref{fig3}a, the resultant total free energy landscape is almost flat but with very small positive curvature. According to Eq.~\ref{capacitance}, the flatter the free energy landscape is, the larger the capacitance is. It can be speculated that the total capacitance is amplified much. The calculated capacitance for 60 nm PZT layer, 1.6 nm MgO and PZT-MgO double layer is shown in Fig.~\ref{fig3}b. It can be seen that the negative capacitance of ferroelectric PZT is not constant while the positive capacitance of dielectric MgO is constant. The total capacitance is much larger than the capacitance of dielectric MgO. Actually, as we calculated, the largest capacitance amplification ratio is 350 times around polarization of $0.15~C/m^2$ and the smallest capacitance amplification ratio is 9 times at polarization of $0.3~C/m^2$.

From Fig.~\ref{fig3}, it is observed that since negative capacitance of PZT is strongly non-linear, the relation $C_{MgO}+C_{FE}\approx 0$ can't always be fulfilled. Even $C_{MgO}$ deviates a little from $C_{FE}$, the voltage amplification ratio will reduce dramatically. The positive capacitance of MgO and the negative capacitance of PZT should be matched to get a appropriate voltage amplification ratio. In order to solve this problem, we have proposed a design methodology which can help us to determine what kind of DE/FE material should be used, how to set the match point between positive capacitance and negative capacitance and the thickness of DE/FE material. The flowchart of the design methodology is sketched in Fig.~\ref{fig3}.

As shown in Fig.~\ref{fig3}, first the measure VCMA coefficient is needed. According to PMA field of the free layer nano-magnet and the VCMA coefficient, the desired electrical field strength which can fully suppress the PMA field of the free layer can be calculated. Also the charge/polarization density of the DE/FE material can be calculated from the electrical field strength. Since $\alpha_1$, $\alpha_{11}$ and $\alpha_{111}$ are material and composition dependent, the free energy landscape and the largest capacitance amplification point in Fig.~\ref{fig3} can be varied to some extent. The appropriate match point which have a large amplification ratio can be set according to  charge/polarization density by varying DE/FE material and FE composition. After that, the DE/FE thickness can be set according to working voltage. If designed DE/FE double layer is not easy for fabrication, the last two steps can be iterated several times.

\begin{figure}
\includegraphics[width=0.45\textwidth]{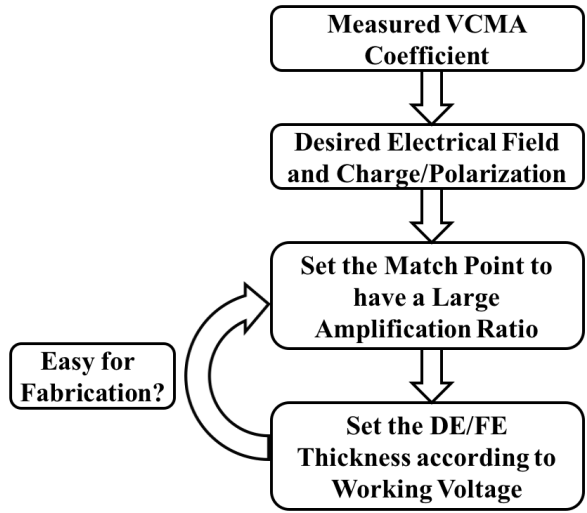}
\caption{\label{fig4} Design methodology to set appropriate DE/FE material, match point and DE/FE thickness.}
\end{figure}

In conclusion, the amplification of VCMA effect by use of negative capacitance is proposed in this work to meet the demand for stronger VCMA effect with scaling MTJ device size. The feasibility of our proposal is demonstrated with physical simulation under Landau theory. In addition, a novel design methodology is also proposed to help determine the DE/FE material and composition, DE/FE thickness and match point.

\begin{acknowledgments}
The authors wish to acknowledge the support from the National Science Foundation of China projects (No. 61306104, 61471015, 61571023, 11201020), the support by the International Collaboration Project 2015DFE12880 from the Ministry of Science and Technology in China and the project from Beijing Municipal of Science and Technology (Grant No. D15110300320000).
\end{acknowledgments}

\bibliography{NC_APL}

\end{document}